\documentclass[12 pt]{article}
\usepackage{times,bm}
\usepackage[affil-it,]{authblk}
\usepackage[pagebackref=false,colorlinks=true,linkcolor=blue,citecolor=blue,urlcolor=blue]{hyperref}
\usepackage{geometry}
\usepackage{graphicx}
\usepackage{cite}
\usepackage{amsmath}
\usepackage{amsfonts}
\usepackage{soul}
\usepackage{amssymb}
\usepackage{subfigure}
\usepackage{color}
\usepackage{float}
\usepackage{multirow,multicol}
\usepackage{tabulary}
\usepackage[flushleft]{threeparttable}
\usepackage{pgfplots,subfigure}
\usepackage{xcolor}
\numberwithin{equation}{section}
\usepackage{tikz}
\usepackage{placeins}
\usepackage{caption}
\usepackage[normalem]{ulem}
\usepackage{hyperref}
\usepackage{nameref}
\usepackage{comment}
\usepackage{pgfplots}
\pgfplotsset{compat=1.7}

\usepgflibrary{arrows}
\usetikzlibrary{shapes.callouts}
\tikzset{
	level/.style   = { thick, },
	connect/.style = { dotted, red   },
	notice/.style  = { draw, rectangle callout, callout relative pointer={#1} },
	label/.style   = { text width=2cm }
}

\usepackage{letltxmacro}
\makeatletter
\let\oldr@@t\r@@t
\def\r@@t#1#2{%
	\setbox0=\hbox{$\oldr@@t#1{#2\,}$}\dimen0=\ht0
	\advance\dimen0-0.2\ht0
	\setbox2=\hbox{\vrule height\ht0 depth -\dimen0}%
	{\box0\lower0.4pt\box2}}
\LetLtxMacro{\oldsqrt}{\sqrt}
\renewcommand*{\sqrt}[2][\ ]{\oldsqrt[#1]{#2}}
\makeatother

\begin{document}
	\newcommand{{\ri}}{{\rm{i}}}
	\newcommand{{\Psibar}}{{\bar{\Psi}}}
	\newcommand{{\red}}{\color{red}}
	\newcommand{{\blue}}{\color{blue}}
	\newcommand{{\green}}{\color{green}}
	\newcommand{\rev}[1]{\textbf{\textcolor{red}{#1}}}
	
	\title{Exploring Non--commutativity as a Perturbation in the Schwarzschild Black Hole: Quasinormal Modes, Scattering, and Shadows}
	
	\author{\large  
		\textit {N. Heidari}$^{\ 1}$ \footnote{E-mail: heidari.n@gmail.com (Corresponding author)},
		\textit {H.Hassanabadi}$^{\ 1, 2}$\footnote{E-mail: hha1349@gmail.com},
		\textit{A. A. Ara\'{u}jo Filho}$^{\ 3}$
		\footnote{dilto@fisica.ufc.br}	and 
		\textit {J. K\u{r}\'\i\u{z}}$^{\ 2 }$\footnote{E-mail: jan.kriz@uhk.cz},
		\\
		\small \textit {$^{\ 1}$Faculty of Physics, Shahrood University of Technology, Shahrood, Iran.}\\
		\small \textit {$^{\ 2}$Department of Physics, University of Hradec Kr$\acute{a}$lov$\acute{e}$, Rokitansk$\acute{e}$ho 62, 500 03 Hradec Kr$\acute{a}$lov$\acute{e}$, Czechia.}\\
		
		\small\textit {$^{\ 3}$ Departamento de Física Teórica and IFIC, Centro Mixto Universidad de Valencia--CSIC. Universidad
			de Valencia, Burjassot-46100, Valencia, Spain}
	}
	
	\date{}
	\maketitle

	\begin{abstract}
		In this work, by a novel approach to studying the scattering of a Schwarzschild black hole, the non--commutativity is introduced as perturbation. We begin by reformulating the Klein--Gordon equation for the scalar field in a new form that takes into account the deformed non--commutative spacetime. Using this formulation, an effective potential for the scattering process is derived.
		To calculate the quasinormal modes, we employ the WKB method and also utilize fitting techniques to investigate the impact of non--commutativity on the scalar quasinormal modes. We thoroughly analyze the results obtained from these different methods. Moreover, the greybody factor and absorption cross section are investigated. 
		Additionally, we explore the behavior of null geodesics in the presence of non--commutativity. Specifically, we examine the photonic, and shadow radius as well as the light trajectories for different non--commutative parameters. Therefore, by addressing these various aspects, we aim to provide a comprehensive understanding of the influence of non--commutativity on the scattering of a Schwarzschild--like black hole and its implications for the behavior of scalar fields and light trajectories.
		
	\end{abstract}
	
	\begin{small}
	\end{small}
	
	\FloatBarrier
	
	%%%%%%%%%%%%%%%%%%%%%%%%%%%%%%%%%%%%%%%%%%%%%%%%%%%%%%%%%%%%%%%%%%%%%%%%%%%%%%%%%%%%%%%%%%%%%%%%%%%%%%%%%%%%%%%%%%%%%%%%%%%%%%%%%%%%%%%%%%%%%%%%%%%%%%%%%%%%%%%%%%%%%%%%%%%%%%%%%%%%%%%%%%%%%%%%%%%%%%%%%%%%%%%%%%%%%%%%%%%%%%%%%%%%%%%%%%%%%%%%%%%%%%%%%%%%%%%%%%%%%%%%%%%%%%%%%%%%%%%%%%%%%%%%%%%%%%%%%%%%%%%%%%%%%%%%%%%%%%%%%%%%%%%%%%%%%%%%%%%%%%%%%%%%%%%%%%%%%%%%%%%%%%%%%%%%%%%%%%%%%%%%%%%%%%%%%%%%%%%%%%%%%%%%%%%%%%%%
	
		\section{Introduction}
Non--commutative spacetime has been a subject of interest for researchers in gravity theories \cite{nicolini2009noncommutative,chamseddine2023noncommutativity}. One significant application of such geometry is certainly in the context of black holes. This spacetime is described by the relation $\left[x^\mu,x^\nu \right]=i \Theta^{\mu \nu}$, where $x^\mu$ represents the operators for spacetime coordinates and $\Theta^{\mu \nu}$ denotes an anti--symmetric constant tensor. Several methods have been developed to incorporate non--commutativity into gravity theories \cite{frob2023noncommutative,nicolini2006noncommutative,modesto2010charged,lopez2006towards,mann2011cosmological}.

Any source, such as the mass (or charge) on the right-hand side of Einstein's equation, has size. The point source used in Schwarzschild's solution is an abstraction commonly used in physics through the Dirac delta function.
Ref. \cite{nicolini2006noncommutative} suggested considering the source mass blurring to implement the related non--commutativity, which is not a direct result of the non-commutativity of spacetime. 
On the other hand, the main effect of non--commutability is on the left side of the gravity equation which affects the space-time measure and causes many different consequences. Regarding this approach, there exists a formalism that involves the use of the non--commutative gauge de Sitter (dS) group, SO(4,1), in conjunction with the Poincarè group, ISO(3,1), regarding the Seiberg--Witten (SW) map  approach\cite{chamseddine2001deforming,araujo2023thermodynamics}. With it, Ref. \cite{chaichian2008corrections} derived a deformed metric for the Schwarzschild black hole space time and recently, Ref. \cite{heidari2023gravitational} introduced a deformed mass approximation corresponds to the deformed metric proposed by 
Ref. \cite{chaichian2008corrections}. As the deformed metric derived from direct complementing non--commutativity space time has been complicated \cite{chaichian2008corrections}, there are different unexplored aspects that we would like to investigate some of them in this work.

The study of gravitational waves and their spectra over the last years has attracted lots of attention, particularly, with the advancements in gravitational wave detectors such as VIRGO and LIGO. 
The interaction of a black hole with its surrounding matter can cause perturbations which are of great importance in gravitational-wave astrophysics and have many applications in black hole physics. The quasinormal modes (QNMs), which are complex oscillation frequencies that appear in the response of black holes to initial perturbation, carry information about the reaction of black holes and various properties such as mass, charge, and angular momentum. Additionally, the imaginary part of QNMs is related to the damping time scale, which can provide insight into the stability of the black hole space-time \cite{grishchuk2001gravitational,rincon2020greybody}.

One aspect of this research involves investigating the quasinormal modes (QNMs), which are complex oscillation frequencies that arise in the response of black holes to initial perturbations. These frequencies can be obtained under specific boundary conditions \cite{berti2009quasinormal,konoplya2011quasinormal}. Various studies have explored the scattering and QNMs in non--commutative spacetime by considering deformed mass density instead of deformed spacetime with a non--commutative (NC) parameter \cite{campos2022quasinormal,panotopoulos2020quasinormal,ciric2018noncommutative,banerjee2008noncommutative,anacleto2020absorption}. 
However, the calculation of quasinormal modes based on a deformed metric has not been extensively addressed in the literature. In this research, we aim to investigate the scattering process of a Schwarzschild black hole in a non--commutative spacetime. To achieve this, we employ the WKB method \cite{iyer1987black,	konoplya2003quasinormal, schutz1985black} to determine the quasinormal frequencies of massless scalar perturbations. Additionally, we calculate the greybody factor for the scalar field and examine the impact of non--commutativity on the absorption cross section.

Furthermore, exploring the geodesics and shadow radius to enhance our understanding of gravitational lensing has gained attention \cite{ovgun20234d,wei2015shadow,anacleto2015gravitational}.
We explore these aspects of the Schwarzschild--like black hole in the non--commutative spacetime.

The structure of this paper is as follows: In Section 2, first we provide an overview of implementing non--commutative spacetime through the metric of the Schwarzschild--like black hole. After that, we focus on the massless Klein--Gordon equation, where we derive a Schr\"{o}dinger--like form for the wave equation and find an effective potential. Section 3 is dedicated to obtaining the quasinormal modes of the non--commutative deformed Schwarzschild black hole using the WKB method, P\"{o}sch--Teller and Rosen--Morse fitting method. In Section 4, we calculate the greybody factor and absorption cross section concerning non--commutativity parameter. Section 5 addresses the null geodesic and shadow radius in non--commutative spacetime. Finally, we present the conclusions in Section 6.

%%%%%%%%%%%%%%%%%%%%%%%%%%%%%%%%%%%%%%%%%%%%%%%%%%%%%%%%%%%

\section{Effective potential of deformed metric by non-commutativity}
In this section, we discuss the non--commutative Schwarzschild black hole spacetime with correction terms. The metric, which takes into account all these features is given by $g_{ij}^{NC}=g_{ij}+\Theta^2 h_{ij}^{NC}$, where $g_{ij}$ represent the original Schwarzschild black hole metric parameters and $h_{ij}^{NC}$s are the coefficients for the non--commutative correction term, as mentioned in Ref. \cite{chaichian2008corrections, chamseddine2001deforming, zet2003desitter}.
Additionally, Ref. \cite{chen2022eikonal} introduces a remarkable proposal for a deformed Schwarzschild black hole metric that is both stationary and axisymmetric. The deformed metric, which incorporates a small dimensionless parameter $\epsilon$, can be expressed in the following form

\begin{align}\label{g00}
	{{\hat g_{00}}} &= - (1 - \frac{{2M}}{r})(1 + \epsilon{A_j}{{\mathop{\rm cos}\nolimits} ^j}\theta ),\\
	{{\hat g}_{11}} &= {(1 - \frac{{2M}}{r})^{ - 1}}(1 + \epsilon{B_j}{{\mathop{\rm cos}\nolimits} ^j}\theta ),\\
	{{\hat g}_{22}}& = {r^2}(1 + \epsilon{C_j}{{\mathop{\rm cos}\nolimits} ^j}\theta ),\\
	{{\hat g}_{33}}&= {r^2}{{\mathop{\rm sin}\nolimits} ^2}\theta (1 + \epsilon{D_j}{{\mathop{\rm cos}\nolimits} ^j}\theta ),\\
	{{\hat g}_{01}}& = \epsilon{a_j}(r){{\mathop{\rm cos}\nolimits} ^j}\theta ,\quad
	{{\hat g}_{12}} = \epsilon{c_j}(r){{\mathop{\rm cos}\nolimits} ^j}\theta ,\quad\
	{{\hat g}_{23}} = \epsilon{e_j}(r){{\mathop{\rm cos}\nolimits} ^j}\theta,\\
	{{\hat g}_{02}}& = \epsilon{b_j}(r){{\mathop{\rm cos}\nolimits} ^j}\theta ,\quad
	{{\hat g}_{13}} = \epsilon{d_j}(r){{\mathop{\rm cos}\nolimits} ^j}\theta. \\ \label{g13}
\end{align}

Influenced by the findings presented in Ref.  \cite{zhao2023quasinormal}, a novel methodology for handling NC spacetime is employed. We approach the NC Schwarzschild metric as a specific instance of the deformed Schwarzschild metric, denoted as $\hat g_{ij}={g_{ij}^{NC}}$. Furthermore, we assume that the small deformed parameter is equivalent to the NC parameter $(\epsilon=\Theta^2)$. Consequently, the deformed coefficients of the metric are derived as indicated in Ref. \cite{chen2022eikonal,zhao2023quasinormal}
\begin{align}\label{A}
	&{A_0} = \frac{{\alpha \left( {8r - 11\alpha } \right)}}{{16{r^3}\left( {r - \alpha } \right)}},
	\\
	&{B_0} =  - \frac{{\alpha \left( {4r - 3\alpha } \right)}}{{16{r^3}\left( {r - \alpha } \right)}},\\
	&{C_0} = \frac{{2{r^2} - 17\alpha \left( {r - \alpha } \right)}}{{32{r^3}\left( {r - \alpha } \right)}},\\
	&{D_0} =  - \frac{{\alpha \left( {2r - \alpha } \right)}}{{16{r^3}\left( {r - \alpha } \right)}},\\
	&{A_j}={B_j}={C_j}=0 \quad \text{and} \quad
	{D_j}= \frac{{1 + {{\left( { - 1} \right)}^j}}}{{32{r^2}}} \quad \text{for}\quad j>0,\\ \label{abcd}
	&{a_j}(r)= {b_j}(r) = {c_j}(r) = {d_j}(r) = 0.
\end{align}

Here, $\alpha$ is a constant given by $\alpha=2M$. Building upon this new approach, the upcoming section focuses on examining the evolution of the massless scalar perturbation field within NC metric, considering the deformed spacetime of the Schwarzschild black hole. To do so, we express the Klein--Gordon equation in the context of the curved spacetime as follows
\begin{equation}\label{klein}
	\Box\psi=\frac{1}{{\sqrt { - g} }}{\partial _\mu }(\sqrt { - g} {g^{\mu \nu }}{\partial _\nu }\psi ) = 0 .
\end{equation}

Assuming two Killing vector $\partial_t$ and $\partial_{\phi}$ the wave function can be decomposed as 
\begin{equation}
	\psi  = \int_{ - \infty }^\infty  {d\omega \sum\limits_{m =  - \infty }^\infty  {{e^{im\varphi }}D_{m,\omega }^2{\psi _{m,\omega }}(r,\theta ){e^{ - i\omega t}}} } ,
\end{equation}
where $D_{m,\omega }^2{\psi _{m,\omega }}(r,\theta )=0$, $m$ and $\omega$ are the azimuthal number and the mode frequency, respectively. Now, if we decompose the operator $D_{m,\omega }^2{\psi _{m,\omega }}$ up to the first order of $\Theta^2$ \cite{chen2022eikonal,	zhao2023quasinormal}, we obtain
\begin{equation}
	D_{m,\omega }^2 = D_{(0)m,\omega }^2 + {\Theta ^2}D_{(1)m,\omega }^2 .
\end{equation}
Applying the metric coefficients from Eq. \eqref{A}- \eqref{abcd} in Eq. \eqref{klein}
\begin{align}
	D_{(0)m,\omega }^2 &=  - ({\omega ^2} - \frac{{{m^2}f}}{{{r^2}{{{\mathop{\rm sin}\nolimits} }^2}\theta }}) - \frac{f}{{{r^2}}}{\partial _r}({r^2}f{\partial _r}) - {{\mathop{\rm cos}\nolimits} ^j}\theta ({\partial _r}({r^2}f{\partial _r}))\\
	&- \frac{f}{{{r^2}{{{\mathop{\rm sin}\nolimits} }^2}\theta }}{\partial _\theta }({\mathop{\rm sin}\nolimits} \theta {\partial _\theta }),\\ \nonumber
	D_{(1)m,\omega }^2 &= \frac{{{m^2}f}}{{{r^2}{{{\mathop{\rm sin}\nolimits} }^2}\theta }}({A_j} - {D_j}){{\mathop{\rm cos}\nolimits} ^j}\theta  - \frac{f}{{{r^2}}}({A_j} - {B_j}){{\mathop{\rm cos}\nolimits} ^j}\theta ({\partial _r}({r^2}f{\partial _r}))\\
	&- \frac{{{f^2}}}{{{r^2}}}({{A'}_j} - {{B'}_j} + {{C'}_j} + {{D'}_j}){{\mathop{\rm cos}\nolimits} ^j}\theta {\partial _r} - \frac{f}{{{r^2}}}({A_j} - {C_j}){{\mathop{\rm cos}\nolimits} ^j}\theta (cot\theta {\partial _\theta } + \partial _\theta ^2)\nonumber \\
	&- \frac{f}{{2{r^2}}}({A_j} + {B_j} - {C_j} + {D_j}){\partial _\theta }{{\mathop{\rm cos}\nolimits} ^j}\theta {\partial _\theta } - \frac{{2i\omega f}}{r}{a_j}{{\mathop{\rm cos}\nolimits} ^j}\theta (r{\partial _r} + 1).\nonumber 
\end{align}
In addition, the tortoise coordinate $r^*$ is proposed as
\begin{equation}
	\frac{{\mathrm{d}r}}{{\mathrm{d}{r^*}}} = f(1 + {\Theta ^2}b_{lm}^j({A_j} - {B_j})).
\end{equation}
Considering that the $\psi _{m,\omega }$ can be expanded with a Legendre functions $P_{lm}(\cos\theta)$ and radial wave functions $R_{l,m}$ as ${\psi _{m,\omega }} = \sum\limits_{l' = \left| m \right|}^\infty  {P_{l'}^m} (\cos\theta){R_{l',m}}(r)$, the radial wave function is related to $\Psi _{m,\omega }$ which satisfies a Schr\"{o}dinger--like equation
\begin{equation}\label{sailm}
	\partial _{{r^*}}^2{\Psi _{lm}} + {\omega ^2}{\Psi _{lm}} = {V_{eff}}(r){\Psi _{lm}}.
\end{equation}
With this expression, the effective potential reads
\begin{equation}\label{Veff}
	%{V_{eff}} = {V_{sch}} +\Theta^2( {V_0} + {V_j})
	{V_{eff}} = {V_{sch}} +\Theta^2 {V_{NC}}.
\end{equation}

Here, $V_{sch}$ is denoted the effective potential in the original form of the Schwarzschild black hole, and $V_{NC}$ is assumed as the NC correction term of the effective potential. After some algebraic manipulations, we explicitly write
\begin{align}
	&{V_{sch}} = f\left( {\frac{{l\left( {l + 1} \right)}}{{{r^2}}} + \frac{{df}}{{dr}}\frac{{\left( {1 - {s^2}} \right)}}{r}} \right),\\
	& V_{NC}= \frac{f}{r}\frac{{df}}{{dr}}b_{lm}^0\left( {{A_0} - {B_0}} \right) + \frac{f}{{{r^2}}}(a_{lm}^0\left( {{A_0} - {D_0}} \right) - c_{lm}^0\left( {{A_0} - {C_0}}\right)\\ \nonumber
	&-\frac{{d_{lm}^0}}{2}({A_0} + {B_0} - {C_0} + {D_0})
	+ \frac{1}{{4{r^2}}}\frac{d}{{d{r^*}}}(b_{lm}^0{r^2}\frac{d}{{d{r^*}}}({A_0} + {B_0} - {C_0} + {D_0})) - \frac{{b_{lm}^0}}{4}\frac{{{d^2}}}{{d{r^*}^2}}({A_0} + {B_0}))\\ \nonumber
	& - \frac{f}{{{r^2}}}\sum\limits_{j = 1}^\infty  {(a_{lm}^j + \frac{1}{2}d_{lm}^j){D_j}}  + \sum\limits_{j = 1}^\infty  {\frac{1}{{4{r^2}}}\frac{d}{{d{r^*}}}(b_{lm}^j{r^2}\frac{d}{{d{r^*}}}){D_j}}. \nonumber
\end{align}

\begin{figure}[ht]
	\centering
	\includegraphics[width=57mm]{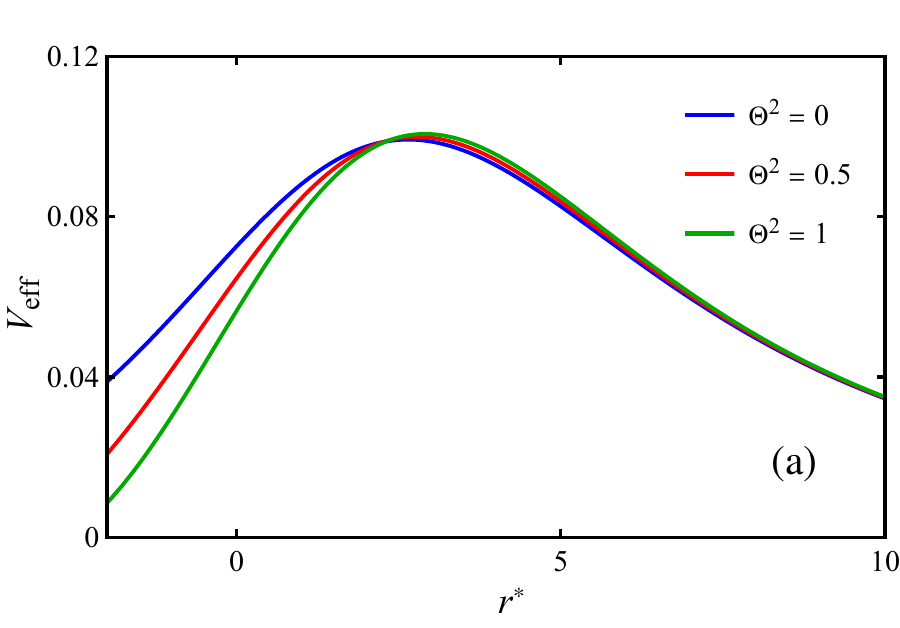} 
	\hfill
	\includegraphics[width=57mm]{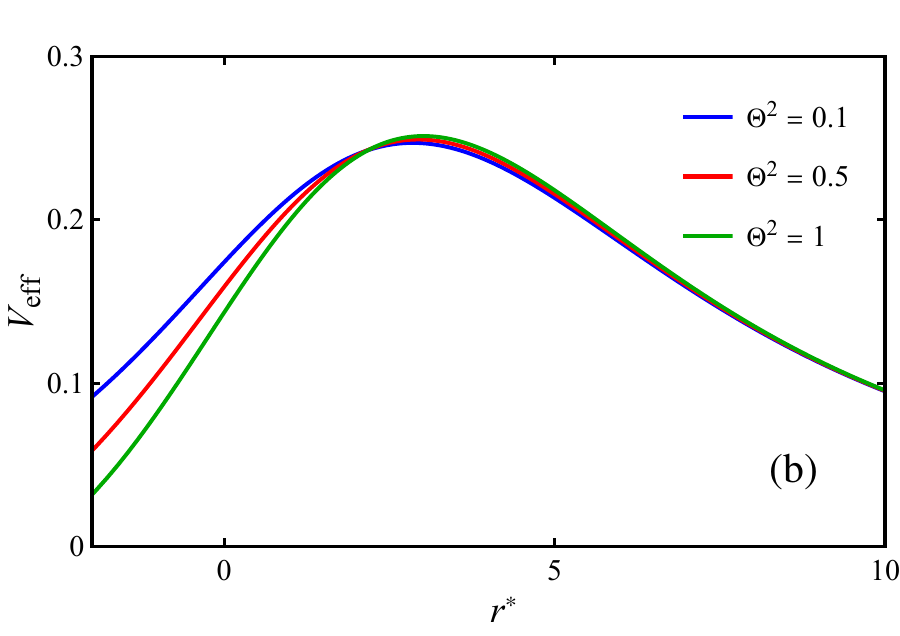}
	\hfil
	\includegraphics[width=57mm]{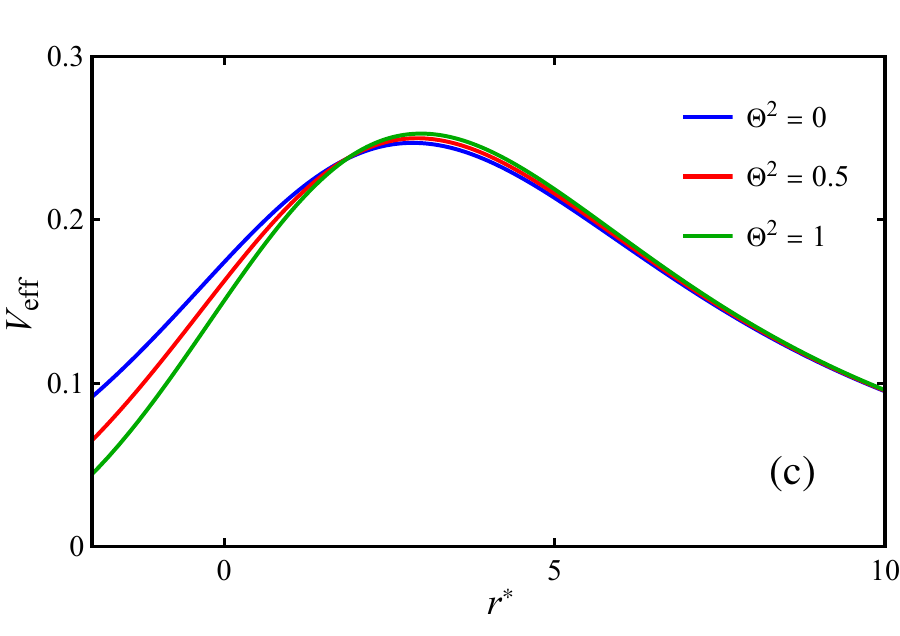}\\
	\caption{Effective potential for scalar field with $M = 1$ in left panel $(a)$ for $l=1  (m=\pm 1) $, in middle panel $(b)$ for $l=2$ $(m=\pm 1)$ and in the right panel $(c)$ for $l=2$ $(m=\pm 2)$ concerning different values of $\Theta^2$. }
	\label{fig:Veffrstar}
\end{figure}

Where $f=1-\frac{2M}{r}$ and the coefficients $a_{lm}^j$, $b_{lm}^j$, $c_{lm}^j$, $d_{lm}^j$ in the effective potential are calculated based on the specific values of the parameters $\Theta$, $l$ and $m$ \cite{zhao2023quasinormal}. It is important to note that the effective potential in a Schwarzschild black hole depends on multipole $l$ and in the presence of NC formalism it also depends on azimuthal number $m$.
We plot the effective potential, denoted as $V_{eff}$, for a given mass $M$, $l$, and $m$ in Fig. \ref{fig:Veffrstar}. In the same figure, we present the effective potential for different values of $\Theta$ when $l=1$ and $m=\pm 1$ in panel (a), and when $l=2$ and $m=\pm 1,2$ in panels (b) and (c), respectively. 

The influence of the NC parameter on the system can be observed through the behavior of the potentials. As the NC parameter increases, the potentials exhibit a higher maximum value. This indicates that the effective potential acts as a stronger barrier to the transmission of the field.
Consequently, the NC effect is expected to have a prominent impact on QNMs and the absorption cross section of the scalar field. The greybody factor, on the other hand, quantifies the probability of transmission through the effective potential barrier.
To explore this further, we can utilize the effective potential to calculate the QNMs and the greybody factor in the following sections.

%%%%%%%%%%%%%%%%%%%%%%%%%%%%%%%%%%%%%%

\section{Non--commutative quasinormal modes}
The QNMs represent the characteristic frequencies at which the scalar field oscillates, and its damping scale. By analyzing the QNMs, we can obtain information about the influence of the NC parameter on the behavior of the scalar field and its interaction with the effective potential. 
The determination of QNMs involves solving the wave equation in Eq. \eqref{sailm} that satisfies specific boundary conditions that require purely incoming waves at the event horizon and purely outgoing waves at infinity. However, due to the complexity of the equation, it cannot be solved analytically. Various analytical and numerical methods \cite{leaver1986solutions,ferrari1984new,heidari2023investigation} have been proposed to find such frequencies. We choose three different methods in the following sections.

\subsection{QNMs with WKB method}
The WKB approximation \cite{iyer1987black,konoplya2003quasinormal} gives the QNM by applying the following formula 
\begin{equation}\label{omegawkb}
	\frac{{i(\omega _n^2 - V_{0})}}{{\sqrt { - 2V''_0} }} + \sum\limits_{j = 2}^3 {{\Omega _j} = n + \frac{1}{2}},
\end{equation} 
Where $V_0$ and $V''_0$ are the value of effective potential and its respective second derivative of the effective potential concerning $r^*$ at its maximum point and $\Omega_j$ are coming from WKB corrections \cite{iyer1987black,
	konoplya2003quasinormal, schutz1985black}.
In Tables. \ref{Table:L1M1}, \ref{Table:L2M1} and \ref{Table:L2M2}, we present the outcome quasinormal frequencies. We consider two families of multipole numbers $l = 1, 2$ and their related monopoles which satisfy $(n \leq l)$, for $M=1$ and different values of $\Theta$. The case when $\Theta = 0$ corresponds to the original Schwarzschild black hole as one should expect.
Our observations reveal that increasing the NC parameter leads to an increase in the real part of the QNMs, indicating a higher propagating frequency. Additionally, the imaginary part of the frequency follows the same trend as the $\Theta$ value increases, suggesting that higher values of NC parameter result in a lower damping timescale for the black hole.

\begin{table}[!ht]
	\centering
	\caption{Quasinormal modes of scalar filed with third order WKB method for $M=1$, $l=1$, $m=1$, $n=0,1$ and various values of $\Theta^2$.}
	\begin{tabular}{|l|l|l|}
		\hline
		$\Theta^2$ & $n=0$ & $n=1$ \\ \hline\hline
		0 & 0.29111 -
		0.09800i & 0.26221 -
		0.30743i \\ \hline
		0.2 & 0.29232 - 0.09940i & 0.26513 -
		0.31105i \\ \hline
		0.4 & 0.29346 - 0.10075i & 0.26754 - 0.31458i \\ \hline
		0.6 & 0.29450 - 0.10199i & 0.26927 - 0.31782i \\ \hline
		0.8 & 0.29532 - 0.10272i & 0.26855 - 0.31928i \\ \hline
		1 & 0.29641 - 0.10434i & 0.27144 - 0.32402i \\ \hline
	\end{tabular}\label{Table:L1M1}
\end{table}

\begin{table}[!ht]
	\centering
	\caption{Quasinormal modes of scalar filed with third order WKB method for $M=1$, $l=2$, $m=1$, $n=0,1,2$ and various values of $\Theta^2$. }
	\begin{tabular}{|l|l|l|l|}
		\hline
		$\Theta^2$ & $n=0$ & $n=1$ & $n=2$ \\ \hline \hline
		0 & 0.48321 - 0.096805i & 0.46319 - 0.29581i & 0.43166 - 0.50343i \\ \hline
		0.2 & 0.48443 - 0.098256i & 0.46522 - 0.30000i & 0.43509 - 0.50993i \\ \hline
		0.4 & 0.48559 - 0.099694i & 0.46699 - 0.30424i & 0.43795 - 0.51669i \\ \hline
		0.6 & 0.48670 - 0.101110i & 0.46849 - 0.30848i & 0.44023 - 0.52356i \\ \hline
		0.8 & 0.48775 - 0.102470i & 0.46963 - 0.31249i & 0.44141 - 0.53004i \\ \hline
		1 & 0.48877 - 0.103800i & 0.47051 - 0.31640i & 0.44188 - 0.53637i \\ \hline
	\end{tabular}\label{Table:L2M1}
\end{table}

\begin{table}[!ht]
	\centering
	\caption{Quasinormal modes of scalar filed with third order WKB method for $M=1$, $l=2$, $m=2$, $n=0,1,2$ and various values of $\Theta^2$. }
	\begin{tabular}{|l|l|l|l|}
		\hline
		$\Theta^2$ & $n=0$ & $n=1$ & $n=2$ \\ \hline\hline
		0 & 0.48321 - 0.096805i & 0.46319 - 0.29581i & 0.43166 - 0.50343i \\ \hline
		0.2 & 0.48465 - 0.098013i & 0.46526 - 0.29937i & 0.43490 - 0.50910i \\ \hline
		0.4 & 0.48606 - 0.099226i & 0.46714 - 0.30300i & 0.43776 - 0.51499i \\ \hline
		0.6 & 0.48741 - 0.100390i & 0.46869 - 0.30642i & 0.43956 - 0.52044i \\ \hline
		0.8 & 0.48873 - 0.101530i & 0.47001 - 0.30976i & 0.44072 - 0.52579i \\ \hline
		1 & 0.49002 - 0.102680i & 0.47121 - 0.31316i & 0.44164 - 0.53133i \\ \hline
	\end{tabular}\label{Table:L2M2}
\end{table}

%%%%%%%%%%%%%%%%%%%%%%%%%%%%%%%%%%%%%%%

\subsection{QNMs with P\"{o}sch--Teller and Rosen--Morse fitting method}

\begin{figure}[ht]
	\centering
	\includegraphics[width=80mm]{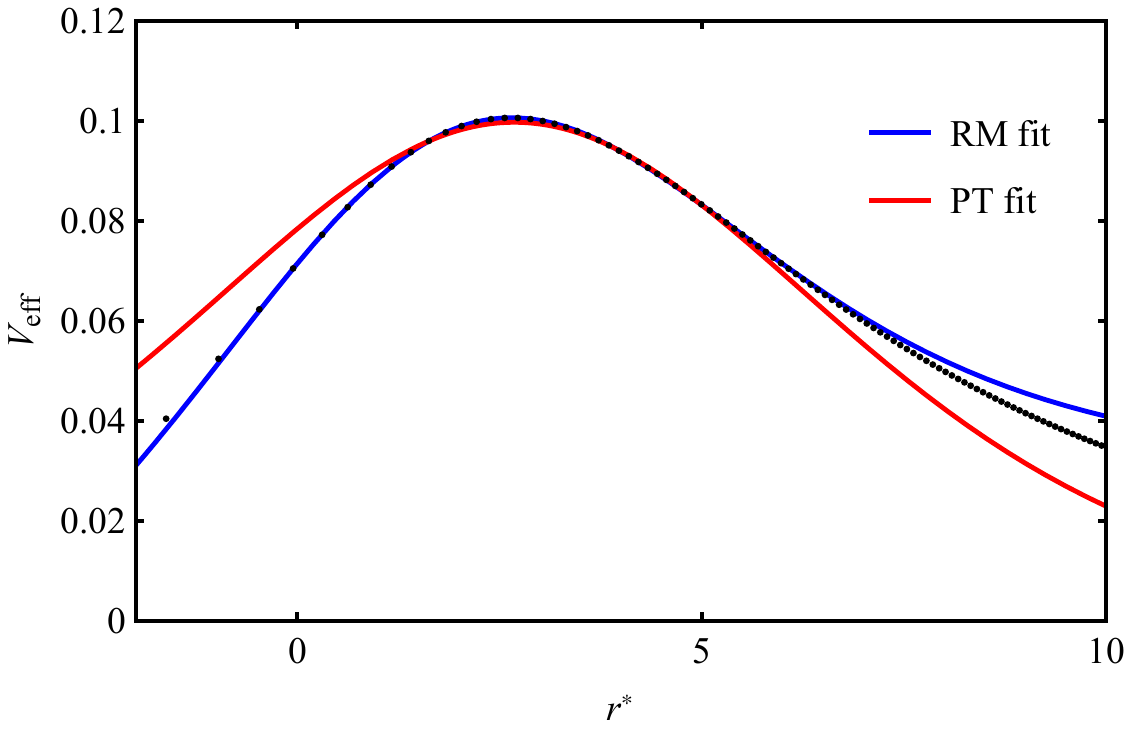}
	\caption{Fitting of $V_{eff}(r^*)$ with P\"{o}sch--Teller and Rosen Morse function for scalar field,  $M=1$, $l=1$ and $\Theta^2=0.5$. Dots show the effective potential points.}
	\label{fig:fitting}
\end{figure}

Another method to solve the Eq. \eqref{sailm} is providing an approximation of the effective potential $V_{eff}$ to a solvable function. We choose the fitting approach to approximate it with the P\"{o}schl--Teller (PT) \cite{ferrari1984new} and Rosen--Morse (RM) functions \cite{heidari2023investigation} for the calculation of QNMs. 	
First, we utilize the \textit{Mathematica} software to fit $V_{eff}$ with PT and RM functions. 
Fig. \ref{fig:fitting} demonstrates the $V_{eff}$ on which the PT and RM functions are fitted, based on the least square method.\\
Next, it is approximated with PT function 
\begin{equation}
	V_{eff}(r^*)\sim V_{PT}=\frac{{{{V_0}}}}{{{{\cosh }^2}\gamma ({r^*} - {{\bar r}^*})}}.
\end{equation}
Here, ${V_0}$ and $\gamma$ are  the height and the curvature of the effective potential $V_{eff}(r^*)$ at its maximum point corresponding to ${\bar{r}}^*$, respectively.
Then, solving Eq. \eqref{sailm} with the P\"{o}schl--Teller approximation leads to the following equation for the calculation of QNMs \cite{ferrari1984new}
\begin{equation}\label{PTQNM}
	\omega = i\gamma (n + \frac{1}{2}) \pm \gamma \sqrt {\frac{{{V_0}}}{{{\gamma ^2}}} - \frac{1}{4}}.
\end{equation}
The coefficients $V_0$ and $\gamma$ are found by fitting program, and the QNMs are calculated according to the Eq. \eqref{PTQNM}.\\
Furthermore, the RM function which has a correction term adding asymmetry to PT, can be considered in solving the Eq. \eqref{sailm} as an approximation of the effective potential 
\begin{equation}
	V_{eff}(r^ *)\sim V_{RM}= 
	\frac{{{{V_{0}}}}}{{{{\cosh }^2}\gamma ({r^*} - {{\bar r}^*})}}+ {V_1}\tanh \gamma ({r^*} - {{\bar r}^*}).
\end{equation}
The solution of the wave function with this approach yields the following expression \cite{heidari2023investigation}
\begin{equation}\label{RMQNM}
	\frac{{\sqrt {{\omega ^2} + {V_1}}  + \sqrt {{\omega ^2} - {V_1}} }}{2} =i\gamma (n + \frac{1}{2}) \pm \gamma \sqrt {\frac{{{{V_{0}}}}}{{{\gamma ^2}}} - \frac{1}{4}}.
\end{equation}
in which $V_1$ is a new parameter added to the PT function for better accuracy. The coefficients $V_{0}$, $V_1$ and $\gamma$ are obtained by fitting method via \textit{Mathematica software}.

All results are represented in Table. \ref{tab:PTRM}. Both the real and imaginary parts of QNMs for multipole numbers $l=1$ and $n=0$, are increasing with higher values of NC parameter.

\begin{table}[]
	\caption{Comparison of QNMs of the scalar field of NC Schwarzschild black hole obtained by using P\"{o}sch-Teller fitting, Rosen--Morse fitting and WKB method for  $M=1$, $l=1$, $m=1$, $n=0$ and various values of parameter $\Theta$.}
	\centering
	\label{tab:PTRM}
	\begin{tabular}{|l|l|l|l|}
		\hline
		$\Theta^2$    & PT fitting method                  & RM fitting method                 & WKB method                \\ \hline\hline
		0        & 0.300322 - 0.090925i & 0.296046 - 0.100235i & 0.29111 - 0.09800i \\ \hline
		0.2 & 0.300972 - 0.091393i  & 0.296157 - 0.101602i & 0.29232 - 0.09940i   \\ \hline
		0.4 & 0.301622 - 0.091872i  & 0.296256 - 0.102960i  & 0.29346 - 0.10075i  \\ \hline
		0.6 & 0.302272 - 0.092345i   & 0.296344 - 0.104310i  & 0.29450 - 0.10199i   \\ \hline
		0.8 & 0.302924 - 0.092809i  & 0.296423 - 0.105652i & 0.29532 - 0.10272i  \\ \hline
		1        & 0.303575 - 0.093290i  & 0.296491 - 0.106990i  & 0.29641 - 0.10434i  \\ \hline
	\end{tabular}
\end{table}
In essence, the results show that both fitting methods and WKB approximation align in the behavior of QNMs in the presence of NC spacetime. 

%%%%%%%%%%%%%%%%%%%%%%%%%%%%%%%%%%%%%%%
\section{Non--commutative Greybody factor and Absorption cross section}
Greybody factors calculation is one of the crucial aspects of scattering issues due to the estimation of the portion of the initial quantum radiation in the vicinity of the event horizon reflected and the amount of radiation that will reach the observer through the potential barrier.
We shall take into account the radial wave Eq. \eqref{sailm} with the boundary conditions for incoming wave and outgoing wave as the following form \cite{iyer1987black,schutz1985black,konoplya2003quasinormal}
\begin{equation}
	\Psi_{\omega l}=
	\begin{cases}
		{{e^{ - i\omega {r^*}}} + \mathrm{R}{e^{i\omega {r^*}}}} \quad \text{, if} \ {r^*} \to -\infty ~ (r \to {r_h})\\
		{\mathrm{T}{e^{ - i\omega {r^*}}}} \quad \quad \quad \quad \text{, if} \ {r^*} \to +\infty ~ (r \to \infty )\\
	\end{cases}
\end{equation}
where $\mathrm{R}$ and $\mathrm{T}$ are the reflection and transmission coefficients, respectively.
The reflection coefficient is obtained by applying 3th order WKB method as \cite{anacleto2021quasinormal,konoplya2020grey} 
\begin{equation}
	|\mathrm{R}|^{2} =  \frac{1}{1+e^{-2i\pi \mathcal{K}}},
\end{equation}
where
\begin{equation}
	\mathcal{K}= \frac{i({\omega}^{2}-V_{0})}{\sqrt{-2 V''_{0}}} - \sum_{j=2}^{6} \Omega_{j}.
\end{equation}
Here $\omega$ is purely real, $V_0$ and $V''_0$ is the effective potential in Eq. \eqref{Veff} and its second derivative at its maximum, $\Omega_{j}$ are coefficients associated with the effective potential.
After finding the reflection coefficient by applying $|\mathrm{T}|^{2}+|\mathrm{R}|^{2} =1$, the transmission coefficient can be calculated \cite{konoplya2020grey,toshmatov2016quasinormal}
\begin{equation}
	|\mathrm{T}|^{2} = \frac{1}{1+e^{+2i\pi \mathcal{K}}}.
\end{equation}

\begin{figure}[ht]
	\centering
	\includegraphics[width=80mm]{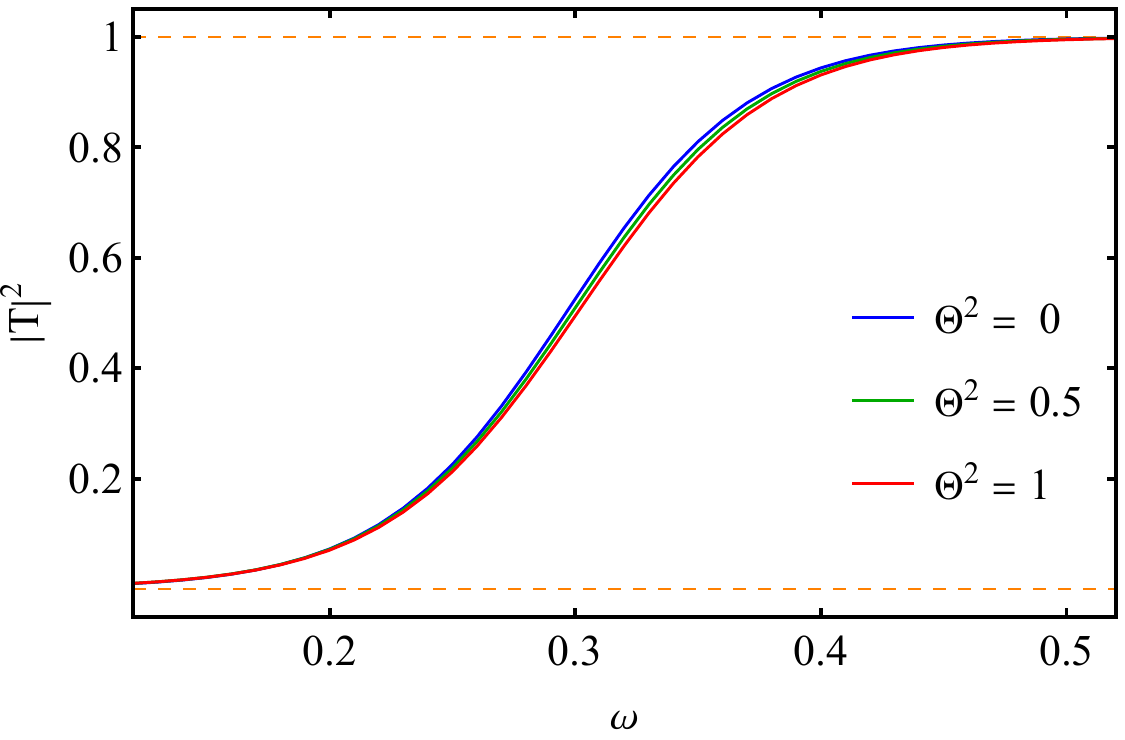}
	\caption{The greybody factors of the scalar field are computed using the third-order WKB method for $M=1$ and $l=1$ and $\Theta^2=0,05,1$. }
	\label{fig:Greyl1}
\end{figure}

As depicted in Fig. \ref{fig:Greyl1}, increasing the value of $\Theta$ leads to a decrease in the greybody factors, indicating that a smaller fraction of the scalar field can penetrate the potential barrier. Additionally, in Fig. \ref{fig:Veffrstar}, it is evident that the height of the potential barrier increases with higher $\Theta$ values, resulting in a lower probability for particles to transmit through the barrier. Consequently, higher values of the NC parameter result in a reduction in the greybody factor and a lower detection of incoming flow by the observer.

To investigate the probability for an outgoing wave to reach infinity, the greybody factors of the scalar field calculated using the third order WKB method are shown in Fig. \ref{fig:Greyl1} for various values of $\Theta$, specifically for multipole $l = 1$.

As the figure shows, increasing the value of $\Theta$ yields a decrease in the greybody factors, indicating a smaller fraction of the scalar field is penetrating the potential barrier. In Fig. \ref{fig:Veffrstar}, for $l=1$ in panel $(a)$ and for $l=2$ in panel $(b)$ and $(c)$, it is obvious that the heights of the potential barriers go up with higher $\Theta$ values, which means the chance of particles to transmit through the barrier becomes lower. Therefore,
higher values of non-commutativity lead to a reduction in the greybody factor and the detection of a lower fraction of incoming flow by the observer.

The partial absorption cross section can be determined by utilizing the transmission coefficient, which is defined as mentioned in Ref. \cite{crispino2009scattering,anacleto2020absorption}

\begin{equation}
	{\sigma^l_{\mathrm{abs}}} = \frac{{\pi (2l + 1)}}{{{{\omega}^2}}}{\left| {{\mathrm{T}_{l}}({\omega} )} \right|^2},
\end{equation}
where $l$ is the mode number and ${\omega}$ is the frequency
\begin{figure}[ht]
	\centering
	\includegraphics[width=80mm]{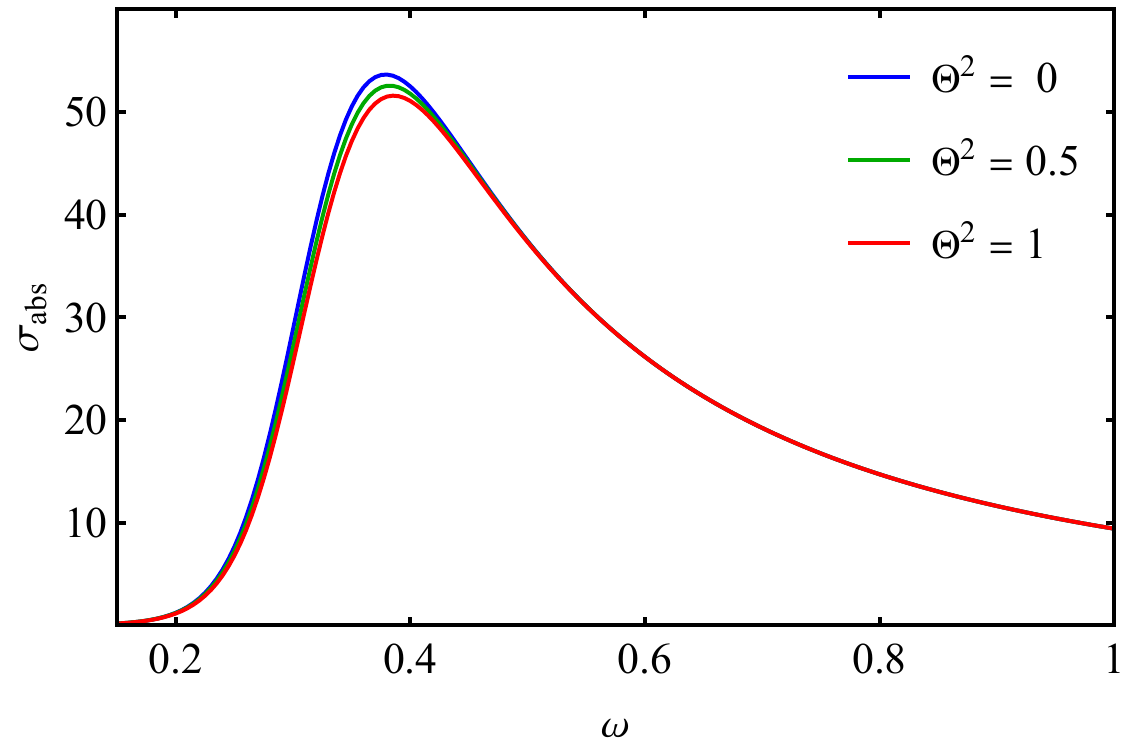}
	\caption{The partial absorption cross section of the scalar field is computed using the third order WKB method for $M=1$ and $l=1$ and various values of $\Theta^2$.}
	\label{fig:cross1}
\end{figure}

In Fig. \ref{fig:cross1}, we have plotted the partial absorption cross section for $\Theta^2 = 0, 0.5, 1 $. As observed, the absorption cross section decreases as the non-commutativity parameter increases. This observation aligns with the fact that the height of the effective potential barrier in Fig.\ref{fig:Veffrstar} increases with the non-commutativity parameter.

\section{Non--commutative Null geodesics, Photonic radius and Shadows}
\begin{figure}[ht]
	\centering
	\includegraphics[width=80mm]{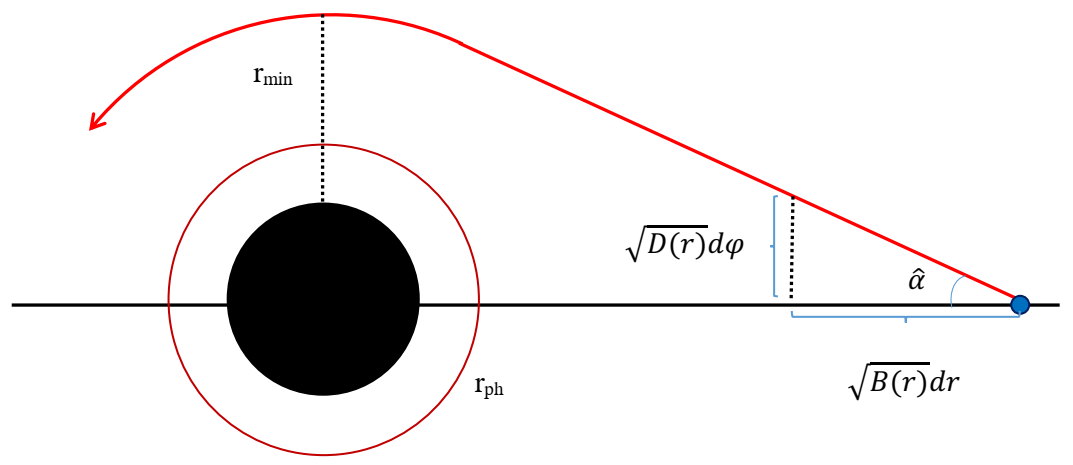}
	\caption{Schematic of light ray emitted from the observer’s position into the past under an angle $\hat\alpha$, $r_{ph}$ is the photon sphere radius and $r_{min}$ denotes the radius of the closest approach.}
	\label{fig:light}
\end{figure}

Another important aspects that worth to be investigating, are the shadows and gravitational lensing in the vicinity of black hole \cite{zeng2022shadows,anacleto2023absorption,yan2023shadows}.  For examination of the black hole shadow radius and null geodesics, similarly what was used in Ref. \cite{batic2015light}, let us consider our diagonal metric, with $g_{ij}^{NC} $ parameters, in the general following form
\begin{equation} 
	{g_{\mu \nu }}\mathrm{d}{x^\mu }\mathrm{d}{x^\nu } =  - A(r)\mathrm{d}{t^2} + B(r)\mathrm{d}{r^2} + C(r,\theta )\mathrm{d}{\theta ^2} + D(r){{\mathop{\rm \sin}\nolimits} ^2}\theta \mathrm{d}{\varphi ^2}.\\
\end{equation}
By applying this form of our metric to the Lagrangian $
L(x,\dot x) = \frac{1}{2}({g_{\mu \nu }}{{\dot x}^\mu }{{\dot x}^\nu }) $, it becomes 
\begin{equation}
	L(x,\dot x) = \frac{1}{2}( - A(r){{\dot t}^2} + B(r){{\dot r}^2} + C(r,\theta ){{\dot \theta }^2} + D(r){{\mathop{\rm \sin}\nolimits} ^2}\theta {{\dot \varphi }^2}).
\end{equation}
Now, we assume the geodesics in the equatorial plane 
$\theta = \frac{\pi }{2}$
which results in $\dot \theta  = 0$ and $
{\mathop{\rm sin}\nolimits} \theta = 1$.
By writing the Euler--Lagrange equation for $t$ and $\phi$, 
\begin{equation}
	\frac{\mathrm{d}}{{\mathrm{d}\tau }}\left(\frac{{\partial L}}{{\partial {{\dot x}^\mu }}}\right) - \frac{{\partial L}}{{\partial {x^\mu }}} = 0 ,
\end{equation}
where we have two constants of motion called $E$ and $L$, which read
\begin{equation}\label{constant}
	E = A(r)\dot t \quad\mathrm{and}\quad L = D(r)\dot \varphi .
\end{equation}
For the sake of convenience, we shall denote $b=\frac{L}{E}$ as being an impact parameter.
For light, ${g_{\mu \nu }}{{\dot x}^\mu }{{\dot x}^\nu } =0$, which means
\begin{equation}\label{light}
	- A(r){{\dot t}^2} + B(r){{\dot r}^2} + D(r){{\dot \varphi }^2} = 0 .
\end{equation}
After applying Eq. \eqref{constant} in Eq. \eqref{light}, the trajectory of light in the equatorial plane can be calculated as 
\begin{equation}\label{rdot}
	\frac{{{{\dot r}^2}}}{{{{\dot \varphi }^2}}} = {\left(\frac{{\mathrm{d}r}}{{\mathrm{d}\varphi }}\right)^2} = \frac{{D(r)}}{{B(r)}}\left(\frac{{D(r)}}{{A(r)}}\frac{{{E^2}}}{{{L^2}}} - 1\right) .
\end{equation}
Following Ref. \cite{perlick2015influence,touati2022geodesic}, we find out the formula for a shadow of an arbitrary spherically symmetric black hole.
When the light ray reaches a minimum radius $r_{min}$ and goes out, $r_{min}$ is assumed as the turning point which satisfies $\mathrm{d}r/\mathrm{d}\varphi =0$. If the function $h(r)$ is proposed as 
\begin{equation}\label{hh0}
	{h^2}(r) = \frac{{D(r)}}{{A(r)}},
\end{equation} 
so that Eq. \eqref{rdot} can be rewritten
\begin{equation}\label{rphi}
	\frac{{\mathrm{d}r}}{{\mathrm{d}\varphi }} =  \pm \frac{{\sqrt {D(r)} }}{{\sqrt {B(r)} }}\sqrt {\frac{{{h^2}(r)}}{{{h^2}(r_{min})}} - 1} .
\end{equation}
In this manner, the turning point has the following relation with the impact parameter 
\begin{equation}
	\frac{{{L^2}}}{{{E^2}}} = \frac{{D(r_{min})}}{{A(r_{min})}} = b^2 \quad \rightarrow \quad h(r_{min}) = b .
\end{equation}
On the other hand, if we call the right hand of the Eq. \eqref{rdot} as ${\tilde V}_{eff}$, we can consider the following equation for the trajectory
\begin{equation}
	{\left(\frac{{\mathrm{d}r}}{{\mathrm{d}\varphi }}\right)^2} + {{\tilde V}_{eff}} = 0 , \quad\quad \text{where}\quad\quad {{\tilde V}_{eff}} =\frac{{ {D(r)} }}{{ {B(r)} }}\left( {\frac{{{h^2}(r)}}{{{h^2}(r_{min})}} - 1} \right).
\end{equation}
Then, the circular orbits can then be determined by solving the expression below
\begin{equation}
	{{\tilde V}_{eff}} = \frac{{\mathrm{d}{{\tilde V}_{eff}}}}{{\mathrm{d}r}} = 0 .
\end{equation}
Using the above conditions, one can determine the radius of the photon sphere $r_{ph}$ by solving
\begin{align}\label{hh}
	&\frac{\mathrm{d}}{{\mathrm{d}r}}{h^2}(r) = 0 \quad \rightarrow \quad {A'(r)D(r) - D'(r)A(r)} = 0.
\end{align}
By taking into account that according to the main metric ($\mathrm{d}s^2=g_{ij}^{NC}\mathrm{d}x^i \mathrm{d}x^j$) in the equatorial plane, we have
\begin{eqnarray}
	A(r)&=&- \left(1 - \frac{{2M}}{r}\right)\left(1 + \frac{{\alpha \left( {8r - 11\alpha } \right)}}{{16{r^3}\left( {r - \alpha } \right)}}{\Theta ^2}\right) ,\\
	D(r)&=&{r^2}\left(1 + \frac{{  - \alpha (2r - \alpha )}}{{16{r^3}(r - \alpha )}}{\Theta ^2}\right).
\end{eqnarray} 
After utilizing the above expressions of $A(r)$ and $D(r)$ in Eq. \eqref{hh}, some algebraic manipulations, the following equation has been found for calculating the photonic radius
	\begin{align}
		&\frac{4 \Theta ^2 M \left(11 M-3 r_{\text{ph}}\right)+8 M r_{\text{ph}}^3}{\Theta ^2 M r_{\text{ph}} \left(4 r_{\text{ph}}-11 M\right)+4 r_{\text{ph}}^4 \left(r_{\text{ph}}-2 M\right)}-\\ \nonumber
		&\frac{\Theta ^2 M \left(2 M^2-2 M r_{\text{ph}}+r_{\text{ph}}^2\right)+8 r_{\text{ph}}^3 \left(r_{\text{ph}}-2 M\right){}^2}{r_{\text{ph}} \left(r_{\text{ph}}-2 M\right) \left(\Theta ^2 M \left(M-r_{\text{ph}}\right)+4 r_{\text{ph}}^3 \left(r_{\text{ph}}-2 M\right)\right)}=0 .
\end{align}
The $r_{ph}$s  are calculated for $M=1$ and various values of $\Theta$ to investigate the influence of non-commutativity on photonic spheres. When $\Theta$ equals zero, the problem reduces to a general Schwarzschild Black hole and we expect $r_{ph}$ and shadow radius to be $3M$ and $3 \sqrt{3}M$, respectively in Table. \ref{Table:shadow}, notably, it is observed that the photonic radius experience decreases when $\Theta$ goes from zero to 1. 
In Fig. \ref{fig:light}, a light ray is shown which is sent from the observer’s position at $ro$ into the past. The light ray angle concerning the radial direction is named $\hat\alpha$ and it satisfies the following relation \cite{perlick2015influence,perlick2022calculating}.
\begin{equation}
	\cot \hat\alpha  = \frac{{\sqrt {{g_{rr}}} }}{{\sqrt {{g_{\varphi \varphi }}} }}\frac{{\mathrm{d}r}}{{\mathrm{d}\varphi }}{|_{r = ro}} = \frac{{\sqrt {B(r)} }}{{\sqrt {D(r)} }}\frac{{\mathrm{d}r}}{{\mathrm{d}\varphi }}{|_{r = ro}}.
\end{equation}
Now by taking into account the Eq. \eqref{rphi} for observer position, we arrive at
\begin{equation}\label{cot}
	{\cot ^2}\hat\alpha  = \frac{{{h^2}(ro)}}{{{h^2}(r_{min})}} - 1, \quad \longrightarrow \quad {\sin ^2}\hat\alpha  = \frac{{{h^2}(r_{min})}}{{{h^2}(ro)}}.
\end{equation}
Therefore, the angular radius of the shadow can be determined by assuming the condition $r_{min} \rightarrow r_{ph} $ in Eq. \eqref{cot}
\begin{equation}
	{\sin ^2}{\hat\alpha _{sh}} = \frac{{{h^2}({r_{ph}})}}{{{h^2}(ro)}}.
\end{equation}
Then, considering the observer at a large distance, the shadow angel can be approximated by
\begin{equation}\label{alpha}
	{\hat\alpha _{sh}} = \frac{{h({r_{ph}})}}{{h(ro)}}.
\end{equation}
On the other hand, $\hat\alpha_{sh}$ approximately has the following relation with the shadow radius.
\begin{equation}\label{alphash}
	{\hat\alpha _{sh}} = \frac{{{R_{sh}}}}{{ro}}.
\end{equation}
Now we compare Eq. \eqref{alpha} and \eqref{alphash} .By considering the observer at infinity we have $h(ro) \longrightarrow ro$, therefore the next equation for shadow radius will be obtained.
	
	\begin{align}\label{shadow}
		&{R_{sh}} = ro\frac{{h({r_{ph}})}}{{h(ro)}} = \sqrt {\frac{{D({r_{ph}})}}{{A({r_{ph}})}}} =\\ \nonumber
		&\sqrt{\frac{\Theta ^2 M r_{\text{ph}}^3 \left(M-r_{\text{ph}}\right)+4 r_{\text{ph}}^6 \left(r_{\text{ph}}-2 M\right)}{\left(2 M-r_{\text{ph}}\right) \left(\Theta ^2 M \left(11 M-4 r_{\text{ph}}\right)+4 r_{\text{ph}}^3 \left(2 M-r_{\text{ph}}\right)\right)}}.
	\end{align}

To investigate the effect of non–commutative parameter on the shadow radius, we applied Eq. \ref{shadow} for various values of the $\Theta$. The results are represented in Table. \ref{Table:shadow}.

\begin{table}[!ht]
	\centering
	\caption{Photonic and shadow radius for $M = 0.5$ and various values of $\Theta^2$}
	\begin{tabular}{|l|l|l|l|l|l|l|}
		\hline 
		$\Theta^2$ & 0 & 0.2 & 0.4 & 0.6 & 0.8 & 1 \\ \hline\hline
		$r_{ph}$ & 1.5 &1.49444 & 1.48884 & 1.48317 & 1.47739 & 1.47147 \\ \hline
		$R_{sh}$ & 2.59808 & 2.5692 & 2.54031 & 2.51136 & 2.48234 & 2.4532 \\ \hline
	\end{tabular}
	\label{Table:shadow}
\end{table}

\begin{figure}[ht]
	\centering
	\includegraphics[width=80mm]{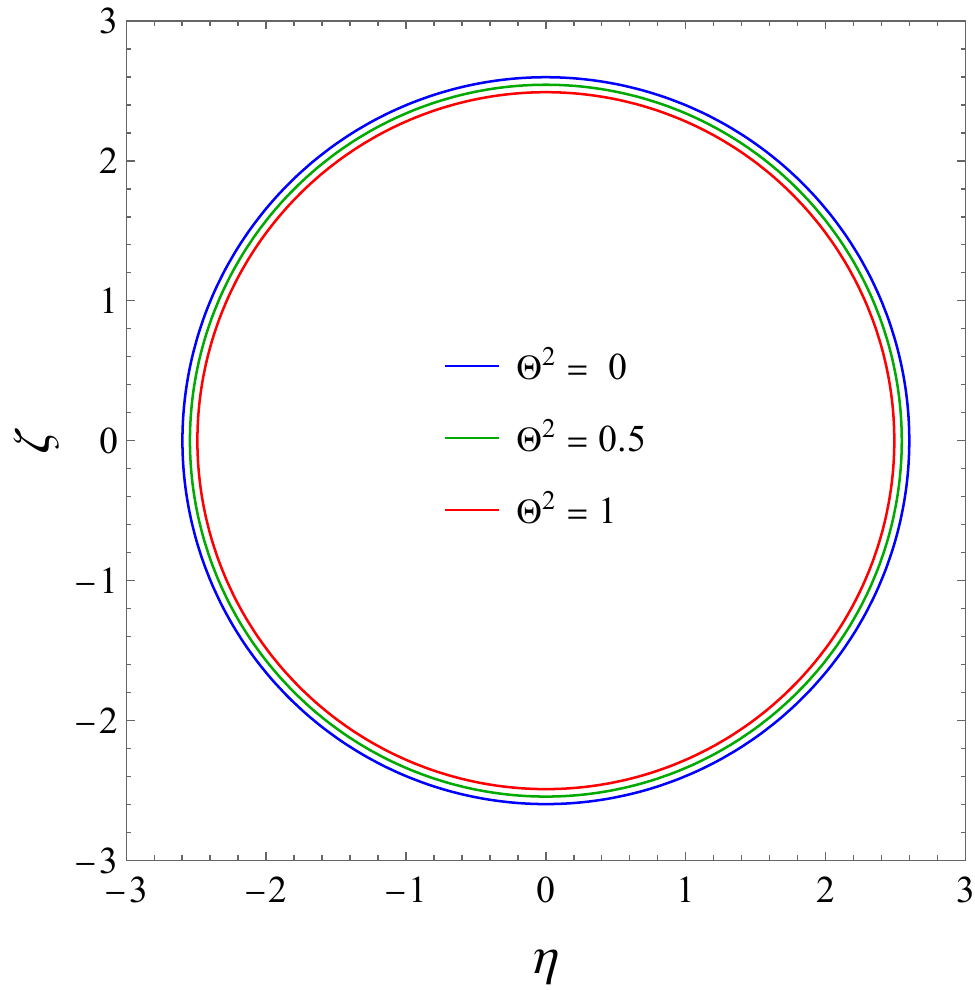}
	\caption{Shadow radius for $M=0.5$ and different NC parameter $\Theta^2=0,0.5,1$ }
	\label{fig:Shadow}
\end{figure}

For better visualization, we display an analysis of the shadows of our black hole for a range of $\Theta$ values with the help of stereographic projection in the celestial coordinates $\eta$ and $\zeta$  \cite{singh2018shadow} in Fig. \ref{fig:Shadow}.  

\begin{figure}[ht]
	\centering
	\includegraphics[width=80mm]{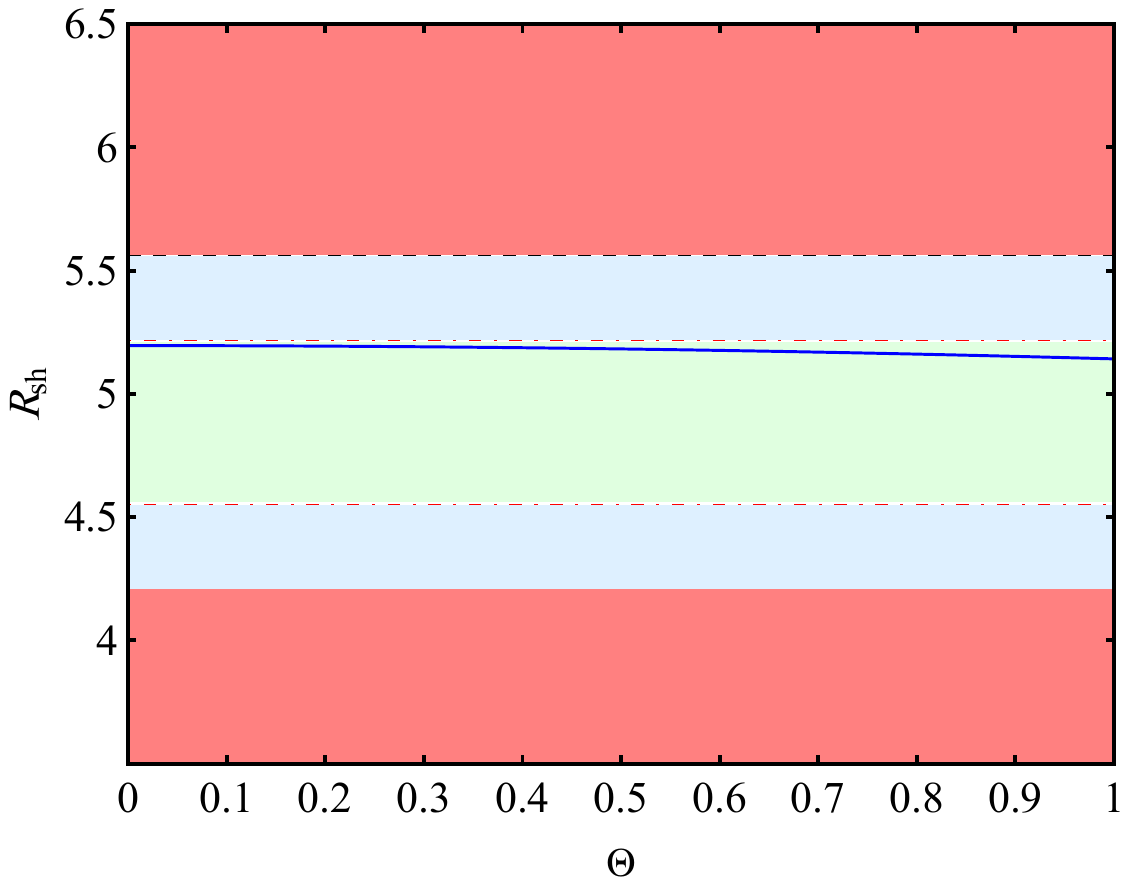}
	\caption{Shadow radius versus $\Theta$ and constraints from EHT horizon-scale image of $Sgr A^*$ at $1\sigma$ and $2\sigma$ }
	\label{fig:constraints}
\end{figure}

The shadow radius demonstrates a reduction as the non-commutative parameter increases, proving that $\Theta$ has a strong effect on the black hole shadow size.\\

\begin{figure}[ht]
	\centering
	\includegraphics[width=55mm]{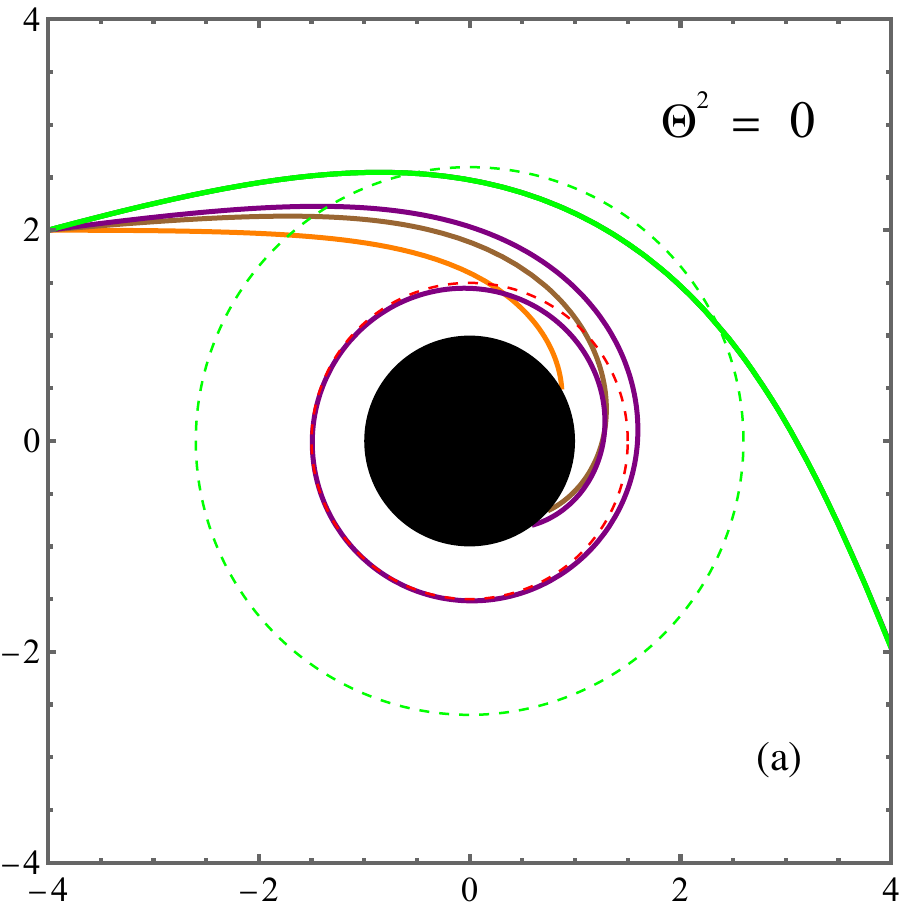} 
	\hfill
	\includegraphics[width=55mm]{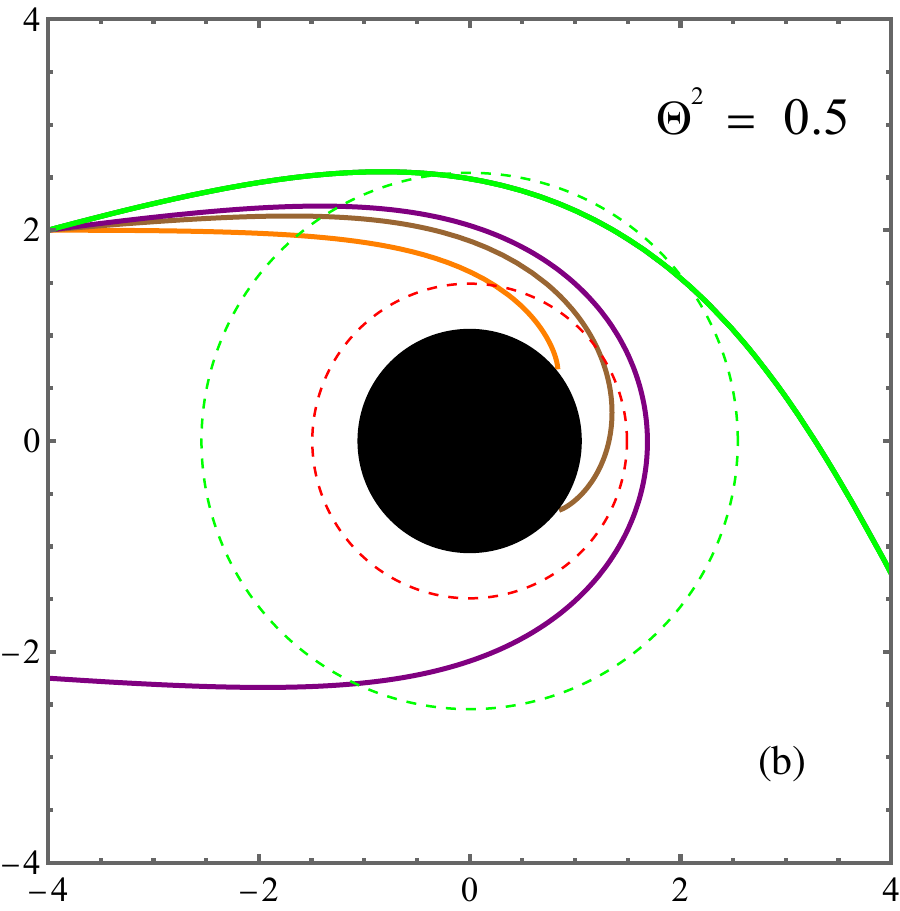}
	\hfil
	\includegraphics[width=55mm]{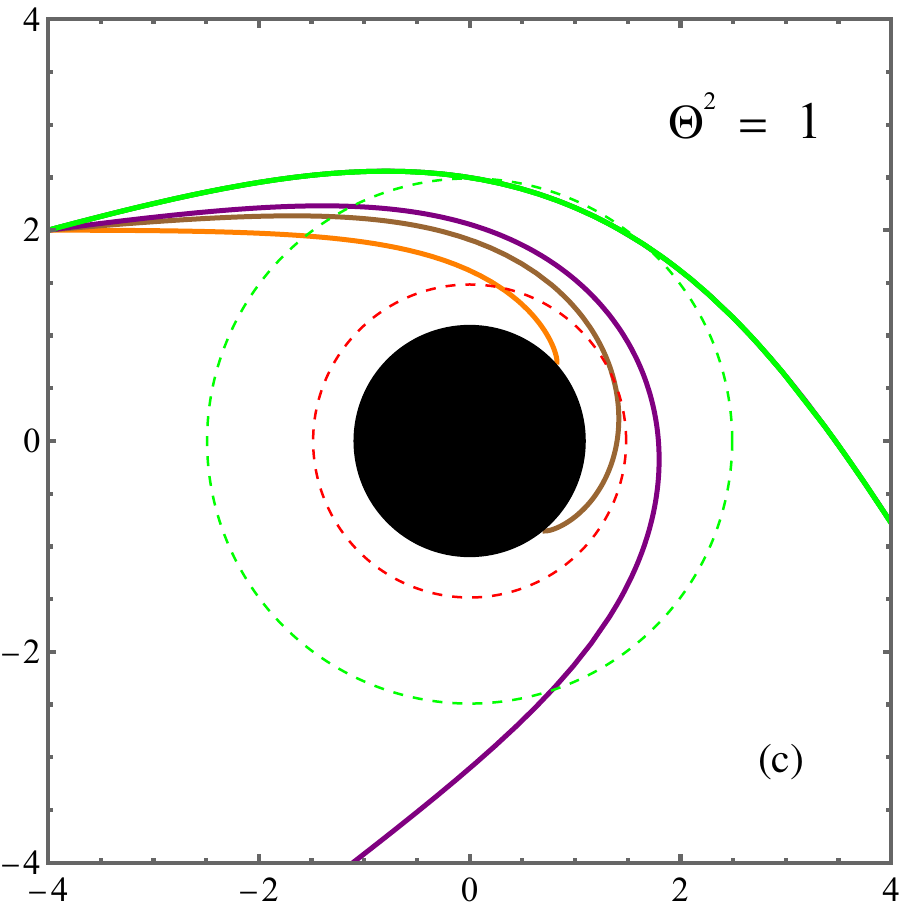}\\
	\caption{Null geodesics representation and in $(a)$ $\Theta^2=0$, in $(b)$ and $(c)$ the light trajectory with $\Theta^2=0.5$ and $\Theta^2=1$, respectively. }
	\label{fig:Geo}
\end{figure}

As the observational data of the EHT collaboration for $Sgr. A ^ *$ places some constraints on the shadow radius \cite{vagnozzi2022horizon,uniyal2023probing}, we can investigate the probable limits for the NC parameter. For this purpose, the shadow radius with respect to $\Theta$ is plotted in Fig. \ref{fig:constraints} for $M = 1$. According to two intervals of $4.55\lesssim R_{sh} \lesssim 5.22$ ($1 \sigma$) and $<4.55\lesssim R_{sh}\lesssim 5.56$ ($2 \sigma$) \cite{vagnozzi2022horizon}, shown with dotted and dashed line in Fig. \ref{fig:constraints}. We can observe that the shadow radius is always consistent with the EHT observations, specifically for $1 \sigma$, which means that the size of $Sgr A^*$ ’s shadow does not place meaningful constraints on the NC parameter. \\

Furthermore, the influence of the non-commutative parameter on the null geodesic curves is in our interest. By solving the Eq. \ref{rdot} numerically, we verify the change of light trajectories for $M = 0.5$ and different values of the non-commutative parameter in Fig. \ref{fig:Geo}. In the figures, we have a black disk that denotes the limit of the event horizon, the internal red dotted circle is the photonic radius, and the external green dotted circle is the shadow radius. Therefore, it is evident that the non-commutative parameter decreases the deflection effect of the black hole on the light beams. The black hole has a weaker influence on the light trajectory for bigger values of $\Theta$.

\section{Conclusion}
In this research, we investigated the influence of non--commutativity as a perturbation in the Schwarzschild black hole. Specifically, we examined the Klein--Gordon equation for a massless scalar field and obtained the effective potential affected by non--commutativity. To calculate the quasinormal mode for certain monopole numbers, we employed three methods. The WKB method and potential approximation using the P\"{o}sch--Teller and Rosen--Morse functions. By analyzing the real and imaginary parts of these frequencies, we obtained a better understanding of how non--commutativity affects the propagation frequency of the scalar field and the damping scale of the black hole.

Our findings indicated that increasing the non--commutativity parameter led to higher scattering frequencies and damping timescales as well. Additionally, as it became stronger, the absorption cross section increased. Furthermore, we calculated the shadow radius, revealing that larger values of $\Theta$ resulted in smaller shadow radii. Lastly, our investigation of null--geodesics suggests that the NC spacetime reduces the gravitational lensing impact of the black hole on the trajectory of light.

%%%%%%%%%%%%%%%%%%%%%%%%%%%%%%%%%%%%%%%%%%%%%%%%%%%%%%%%%%%%%%%%%%%%%%%%%%%%%%%%%%%%%%%%%%%%%%%%%%%%%%%%%%%%%%%%%%%%%%%%%%%%%%%%%%%%%%%%%%%%%%%%%%%%%%%%%%%%%%%%%%%%%%%%%%%%%%%%%%%%%%%%%%%%%%%%%%%%%%%%%%%%%%%%%%%%%%%%%%%%%%%%%%%%%%%%%%%%%%%%%%%%%%%%%%%%%%%%%%%%%%%%%%%%%%%%%%%%%%%%%%

%%%%%%%%%%%%%%%%%%%%%%%%%%%%%%%%%%%%%%%%%%%%%%%%

% Acknowledgement

%%%%%%%%%%%%%%%%%%%%%%%%%%%%%%%%%%%%%%%%%%%%%%%%

\section*{Acknowledgements}

	The authors are grateful to the referees for the invaluable feedback provided to enhance the manuscript's quality. Moreover, the research was partially supported by the Long–Term Conceptual Development of a University of Hradec Kralove for 2023, issued by the Ministry of Education, Youth, and Sports of the Czech Republic. 
%% REFERENCES

%%%%%%%%%%%%%%%%%%%%%%%%%%%%%%%%%%%%%%%%%%%%%%%%
%%%%%%%%%%%%%%%%%%%%%%%%%%%%%%%%%%%%%%%%%%%%%%%%%%%%%%%%%%%%%%%%%%%%%%%%%%%%%%%%%%%%%%%%%%
\section{Data Availability Statement}

Data Availability Statement: No Data associated with the manuscript

%%%%%%%%%%%%%%%%%%%%%%%%%%%%%%%%%%%%%%%%%%%%%%%%%%%%%%%%%%%%%%%%%%%%%%%%%%%%%%%%%%%%%%%%%%
\bibliographystyle{ieeetr}
\bibliography{main}

\end{document}